\documentclass[aps,prl,showpacs,floatfix,twocolumn,superscriptaddress]{revtex4-1}
\usepackage{graphicx}    
\usepackage{hyperref}    
\usepackage{bm}          	
\usepackage[sumlimits,intlimits]{amsmath}
\usepackage{amsfonts,amssymb}
\usepackage{tikz}
\usepackage{cancel}
\usetikzlibrary{arrows, decorations.pathmorphing}
\pdfoutput=1

\begin{document}

\title{Strain-induced gap modification in black phosphorus}

\author{A. S. Rodin}
\affiliation{Boston University, 590 Commonwealth Ave., Boston MA 02215}
\author{A. Carvalho}
\affiliation{Graphene Research Centre and Department of Physics, National University of Singapore, 117542, Singapore}
\author{A. H. Castro Neto}
\affiliation{Boston University, 590 Commonwealth Ave., Boston MA 02215}
\affiliation{Graphene Research Centre and Department of Physics, National University of Singapore, 117542, Singapore}

\date{\today}
\begin{abstract}
The band structure of single-layer black phosphorus and the effect of strain 
are predicted using density functional theory and tight-binding models.
Having determined the localized orbital composition of the individual bands from first-principles, 
we use the system symmetry to write down the effective low-energy Hamiltonian at the $\Gamma$ point. 
From  numerical calculations  and  arguments based on the crystal structure of the material,
we show that the deformation in the direction normal to the plane can be use to change the gap size
and induce a semiconductor-metal transition.

\end{abstract}


\maketitle

\emph{Introduction---}
Despite the fact that the variety of truly two-dimensional materials has been increasing rapidly in the recent years~\cite{Alem2009ath,Lalmi2010ego, Wang2012eao}, graphene is set apart from the rest  because it contains a single non-metal atom type. In fact, the choices for such monotypic systems composed of light non-metals are limited. Carbon is the only solid non-metal in the second period of the periodic table. The third period contains two such elements: phosphorus and sulfur. Phosphorus is a pnictogen and, as such, typically forms three bonds. This means that it is possible to generate a plane of phosphorus atoms, where every atom has three neighbors. Indeed, there exists a phosphorus allotrope, known as black phosphorus, in which atoms form two-dimensional layers. The layers are held together by weak van der Waals force, similarly to graphene. There are two main traits that set black phosphorus apart from the famous carbon allotrope. First, since P atoms are substantially heavier than C, one expects that spin-orbit interaction in phosphorus materials will be stronger. On the structural side, unlike graphene the layers are not perfectly flat; instead, they form a puckered surface due to the $sp^3$ hybridization.

Previous work dealing with black phosphorus monolayers focused on obtaining the band structure using extended tight-binding modeling~\cite{Takao1981eso} and \emph{ab initio} calculations~\cite{Du2010ais}. In this paper, we employ the first principles calculation in order to construct an effective low-energy Hamiltonian. 
Further, we show that uniaxial stress along the direction perpendicular to the layer
can be used to change the gap size in the system,
transforming the material into a 2D metal.

\emph{Structure---}We begin our discussion by looking at the structure of black phosphorus. As a start, it is helpful to consider the best known phosphorus allotrope: white phosphorus. It is described by the molecular formula $\text{P}_4$. The atoms in the molecule form a tetrahedron with six single bonds so that every P atom has three bonds with its neighbors. From the valence shell electron pair repulsion (VSEPR) theory, one can determine that each atom also has a single lone pair. Three bonds and a lone pair result in the $sp^3$ hybridization of the 3$s$ and 3$p$ atomic orbitals. Typically, for such a hybridization, bonds and lone pairs stemming from an atom form angles of about $109.5^\circ$. However, because of the molecular structure of $\text{P}_4$, the angles between the bonds are $60^\circ$. This small angle results in a strain that gives rise to the well-known instability of white phosphorus.~\cite{ChemOfEl}

\begin{figure}
\includegraphics[width = 2.5in]{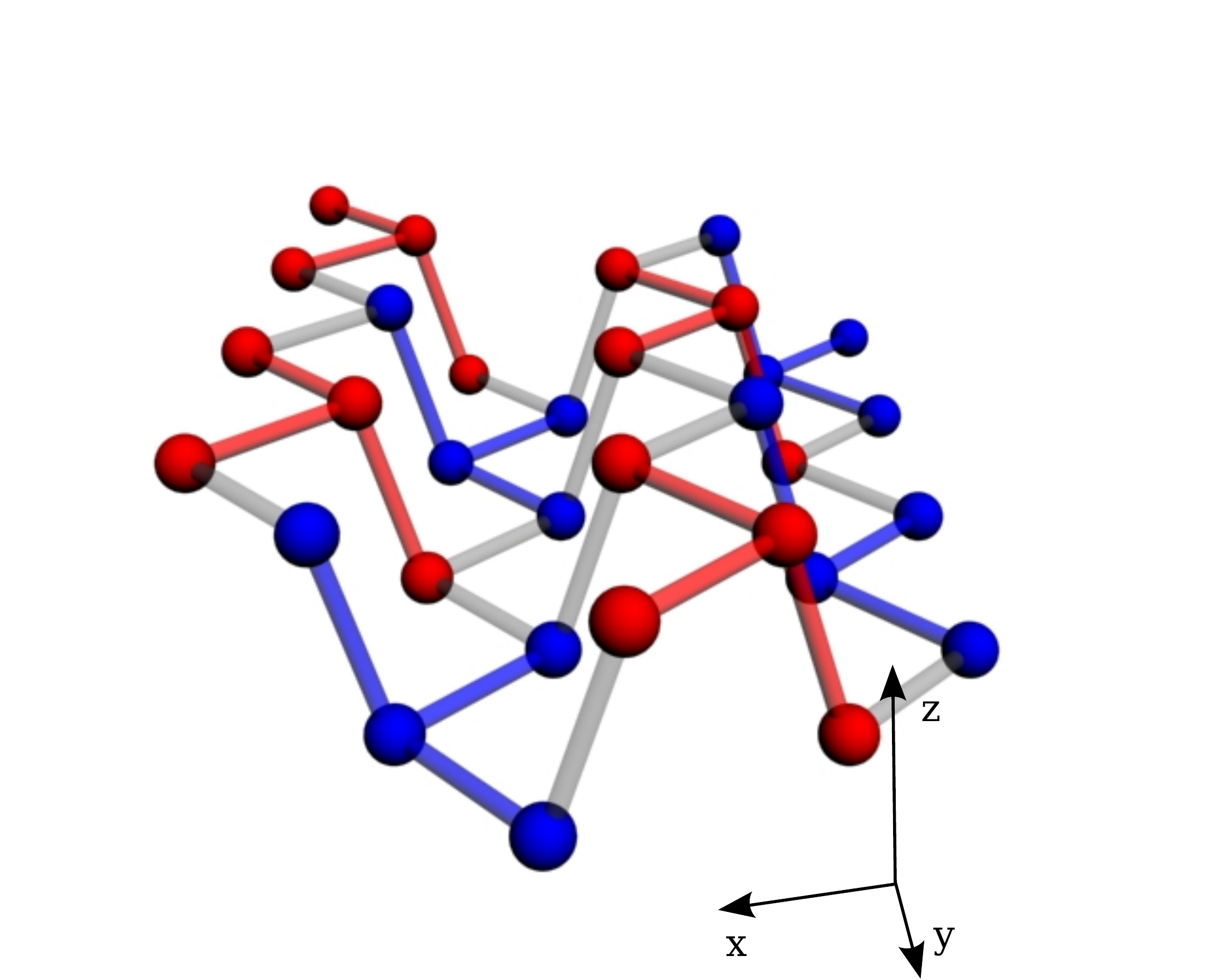}
\caption{(Color online) P-black monolayer lattice structure in three dimensions. The colors represent two different orientations of the flattened $\text{P}_4$ clusters forming the layer. All bonds are identical and the colors are used only as a guide to make the symmetry more apparent.}
\label{fig:Lattice}
\end{figure}

Subjecting white phosphorus to high pressure yields black phosphorus. In this case, three out of six bonds in $\text{P}_4$ become broken, resulting in a ``tripod"-like shape. Because of the bond breaking, the angles between the remaining bonds become larger, making black phosphorus the most stable allotrope of the element. These flattened $\text{P}_4$ blocks form the black phosphorus layer by having their single-bonded atoms link up with two atoms from other blocks. Despite the partial flattening of the four-atom P clusters, they still retain the $sp^3$ hybridization character of tetraphosphorus. Because of this, when linked together, the clusters do not form a flat layer and instead result in a puckered structure, see Fig.~\ref{fig:Lattice}. The illustration shows that the layer is composed of two different orientations of flattened $\text{P}_4$ structures, denoted by the two colors. These two orientations are related by a $180^\circ$ rotation around the $y$-axis, which runs parallel to the direction of puckering steps. A single buckle is made up of alternating $\text{P}_4$ components. The rest of the lattice is generated by replicating these single steps in $x$-direction.

\begin{figure}
\includegraphics[width = 2.5in]{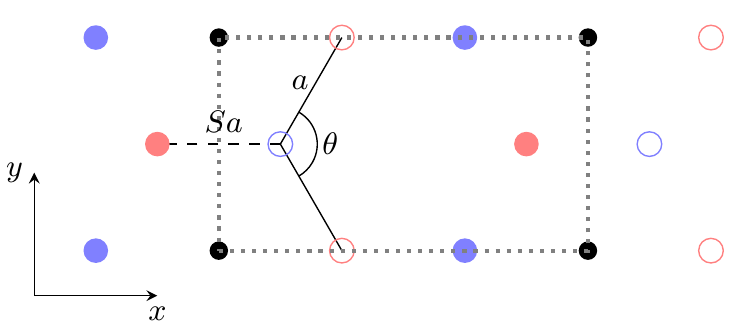}
\caption{(Color online) Projection of the P-black lattice onto $x$-$y$ plane. Filled circles correspond to atoms at the top of the buckles, empty circles are at the bottom. Different colors represent different sublattices. Solid lines are in-plane bonds; dashed one is out-of-plane. The length of in-plane bonds is $a$ and the length of the projection of the out-of-plane ones onto the plane is $Sa$. The unit cell is delineated by the dotted line.}
\label{fig:Projection}
\end{figure}

Unlike flat graphene, characterized by in-plane $\sigma$ and out-of-plane $\pi$ bonds, hybridization in black phosphorus results in orbitals that are composed of $s$ and $p$ components. In addition, the puckering breaks the reflection symmetry in $z$ and $x$ directions. This means that while graphene is described by the $D_{6h}$ point group, black phosphorus has the $C_{2h}$ symmetry with the principal axis running along the puckering steps. Finally, also because of the puckering, a unit cell now contains four atoms: $C^T$, $D^T$, $C^B$, $D^B$ (see Fig.~\ref{fig:Projection}). Here, $C$ and $D$ denote the sublattice; $T$ and $B$ label the top and bottom of the steps.

\emph{First-principles calculation---}
We use first-principles calculations based on density-functional theory
to obtain the bandstructure of monolayer black phosphorus.
These were performed using the {\sc Quantum ESPRESSO} code.\cite{Giannozzi2009}
The core electrons were treated using the projector augmented wave method.\cite{blochl-PhysRevB.50.17953}
The exchange correlation energy was described by the generalized gradient approximation (GGA)
using the PBEsol functional.\cite{PhysRevLett.100.136406}
Since the order of the conduction bands is very sensitive to strain, 
this functional was chosen to obtain the accurate structural parameters (see Supplemental Material).
The Kohn-Sham orbitals were expanded in a plane-wave basis with a cutoff energy of 70~Ry.
The Brillouin-zone (BZ) was sampled using 10$\times$8 points following the scheme proposed by Monkhorst-Pack\cite{Monkhorst1976spf}.

\begin{figure}
\includegraphics[width = 3in]{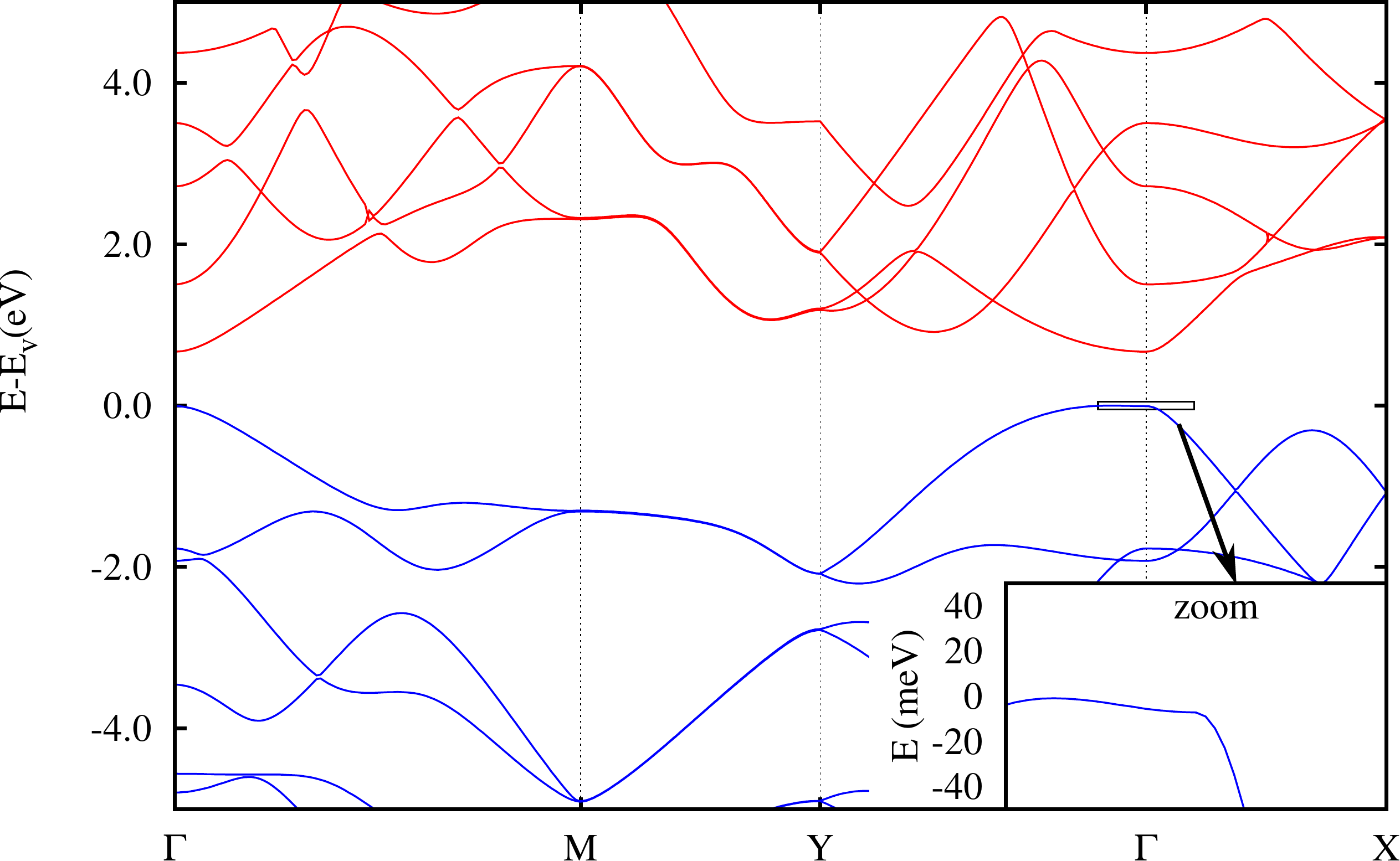}
\caption{First-principles band structure of monolayer black phosphorus.}
\label{fig:Bands}
\end{figure}

Black phosphorus monolayer is a direct bandgap or nearly-direct bandgap semiconductor (Fig.~\ref{fig:Bands}).
The bottom of the conduction band is at $\Gamma$ (Fig.~\ref{fig:Bands}).
The valence band top is also close to the $\Gamma$ point,
and it is nearly dispersionless along the $y$ direction.
The first-principles calculations place its maximum less than 0.06$\times2\pi/a_2$ away from $\Gamma$
, where $a_2$ is the lattice parameter along the $y$ direction (see Supplementary Information). 
The bandgap energy obtained by density functional theory at the GGA level is 0.8~eV.

Similar to the bulk material,
the top of the valence band has predominantly $p_z$ character 
while the lowest conduction bands at $\Gamma$ have mixed $p_x$ and $p_z$ character.
The conduction band which increases in energy in the direction $X$ to $\Gamma$,
i.e. the fourth lowest unoccupied band, has $p_y$ character.
All those four conduction bands are very close in energy and, 
as discussed in Supplemental Material, their relative energy order 
is very sensitive to deformation along the $x$ direction.

\emph{Low-energy Hamiltonian---}Having obtained the band structure using \emph{ab initio} calculations, we now construct a simplified model that describes the bands around the $\Gamma$ point. 
Since the valence band maximum is very close to $\Gamma$ both in the reciprocal space and in energy,
we assume the approximation that the bandgap is direct.
To construct the model, we employ the $\mathbf{k}\cdot \mathbf{p}$ approximation. In this case, the perturbing Hamiltonian is given by $H_1 =\hbar(k_x\hat p_x+k_y\hat p_y)/(2m_e)$. The true eigenstates of the system at the $\Gamma$ point are either even or odd with respect to $\sigma_h$ reflection and can written as sums over irreducible representations of the $C_{2h}$ point group:
\begin{equation}
|\Psi_i^e\rangle = |A_g^i\rangle+|B_u^i\rangle\,,
\quad
|\Psi_i^o\rangle = |A_u^i\rangle+|B_g^i\rangle\,,
\end{equation}
where $A_{u/g}$ and $B_{u/g}$ are the irreducible representations. Using the symmetry argument, we show how the different bands mix through the perturbing Hamiltonian by rewriting the matrix element $\langle \Psi_i^s|\hat p_{x/y}|\Psi_{j}^{s'}\rangle$ as
\begin{equation}
\langle \Psi_i^s|\sigma_h^\dagger\sigma_h\hat p_{x/y}\sigma_h^\dagger\sigma_h|\Psi_{j}^{s'}\rangle = \pm ss'\langle \Psi_i^s|\hat p_{x/y}|\Psi_{j}^{s'}\rangle\,,
\end{equation}
where $s,\, s'=\pm1$ are the $\sigma_h$ symmetry indices. This result tells us that the matrix element for $\hat p_x$ ($\hat p_y$) is nonzero only if the states have the same (different) $\sigma_h$-symmetry.

According to the first principles calculations, the valence and the conduction bands are even in $\sigma_h$. Thus, to the lowest order, the effective low-energy Hamiltonian is
\begin{equation}
H_\text{eff}^0 = \begin{pmatrix}
E_c&\gamma_1k_x
\\
\gamma_1^*k_x&E_v
\end{pmatrix}\,,
\label{eqn:Heff0}
\end{equation}
where $\gamma_1 = \hbar\langle \Psi_c|\hat p_x|\Psi_v\rangle/(2m_e)$. Note that without including the rest of the bands, Eq.~\eqref{eqn:Heff0} describes a one-dimensional system. The lack of $y$-dependence agrees by the weak dispersion in the $y$-direction close to the $\Gamma$ point seen in the numerical results. The rest of the $H_\text{eff}$ is obtained by including the remaining bands and using the L\"{o}wdin partitioning.~\cite{Winkler} Since the principal axis lies in plane of the material, none of the states decouple from others unlike the $p_z$ states in graphene. The leading order correction to the effective Hamiltonian is given by
\begin{equation}
\left(H_\text{corr}\right)_{mm'} = \sum_l\frac{\left(H_1\right)_{ml}\left(H_1\right)_{lm'}}{2}\left[\frac{1}{E_m-E_l}+\frac{1}{E_{m'}-E_l}\right]\,,
\end{equation}
where the summation goes over the remaining bands. The diagonal elements of the correction are
\begin{equation}
\left(H_\text{corr}\right)_{mm} = \sum_l\frac{\left(\gamma_{ml}^x\right)^2k_x^2+\left(\gamma_{ml}^y\right)^2k_y^2}{E_m-E_l} = \eta_mk_x^2+\nu_mk_y^2\,.
\end{equation}
This result captures the mass difference between the conduction and the valence bands, as well as the $\hat x$ and $\hat y$ directions.

\begin{figure}
\includegraphics[width = 3in]{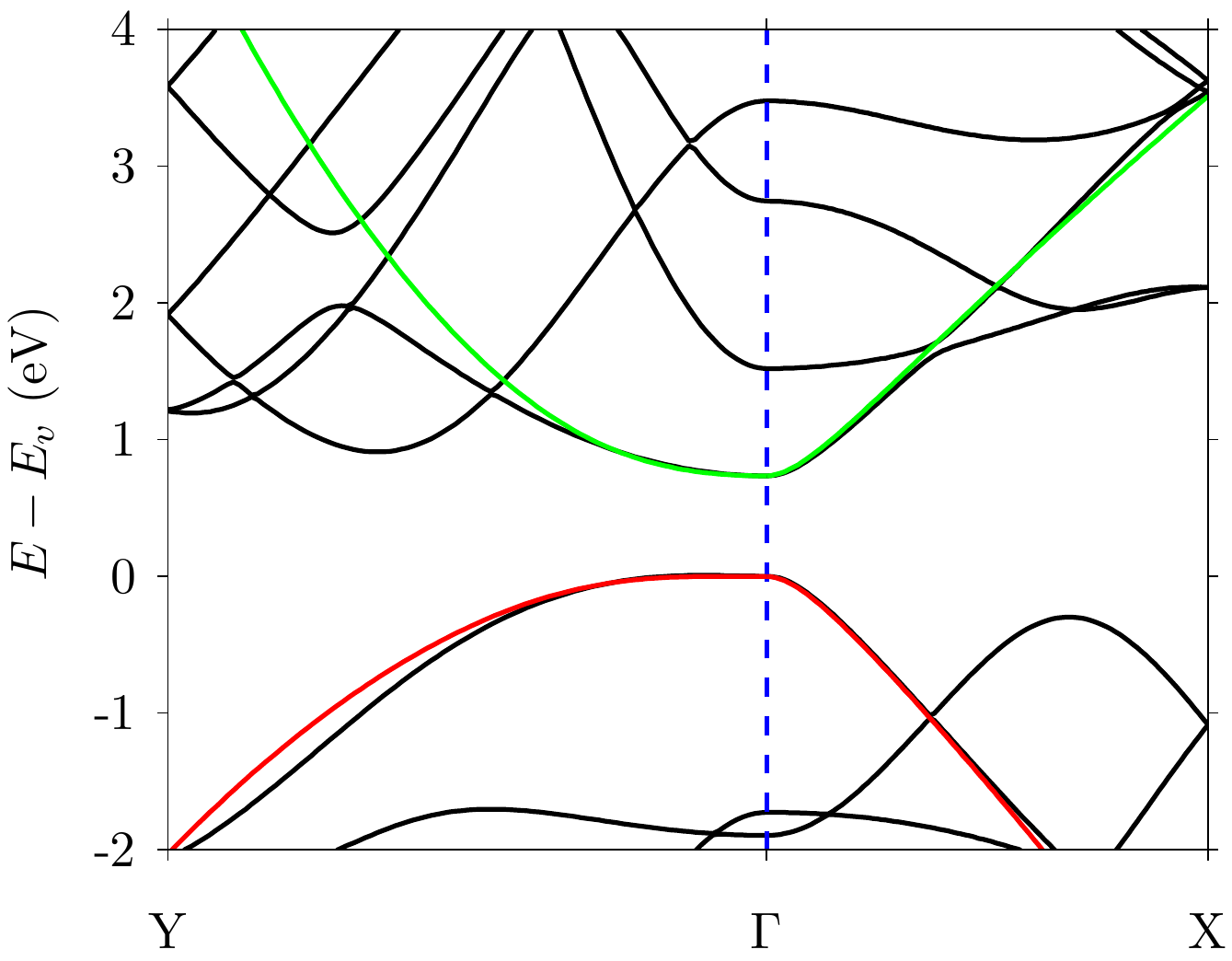}
\caption{Fitting of the $H_\text{eff}$ results to the \emph{ab initio} band structure. For this fit, $\gamma_1 = 6.85$, $\eta_v = -3$, $\nu_c = 5$, and $\beta = 7$. The rest of the parameters are set to zero. Note that this fit gives a direct gap.}
\label{fig:Fit}
\end{figure}

Finally, the off-diagonal elements are
\begin{align}
&(H_\text{corr})_{cv} = \alpha k_x^2+\beta k_y^2\,,
\\
&\alpha = \sum_{l,\text{even}\,\sigma_h}\frac{\gamma^x_{cl}\gamma^x_{vl}}{2}\left[\frac{1}{E_c-E_l}+\frac{1}{E_{v}-E_l}\right]\,,
\\
&\beta = \sum_{l,\text{odd}\,\sigma_h}\frac{\gamma^y_{cl}\gamma^y_{vl}}{2}\left[\frac{1}{E_c-E_l}+\frac{1}{E_{v}-E_l}\right]\,,
\end{align}
resulting in
\begin{equation}
H_\text{eff} = \begin{pmatrix}
E_c+\eta_ck_x^2+\nu_ck_y^2&\gamma_1k_x+\alpha k_x^2+\beta k_y^2
\\
\gamma_1^*k_x+\alpha k_x^2+\beta k_y^2&E_v+\eta_vk_x^2+\nu_vk_y^2
\end{pmatrix}\,.
\end{equation}
From this, one can obtain the effective masses close to the $\Gamma$ point:
\begin{align}
m^x_{c/v}= \frac{\hbar^2}{2\left(\pm\frac{|\gamma_1|^2}{\Delta}+\eta_{c/v}\right)}\,,\quad
m^y_{c/v} = \frac{\hbar^2}{2\nu_{c/v}}\,.
\end{align}
Close to the $\Gamma$ point, we retain only the leading coupling terms and set $\alpha = 0$. We plot a fit for the conduction and valence bands in Fig.~\ref{fig:Fit}.

\emph{Lattice deformation---}
Application of uniaxial stress in the direction perpendicular to the monolayer modifies the band structure. It is known that the symmetry breaking in general lifts degeneracies and opens gaps. In this system, however, vertical compression does not break symmetry. Rather, as the thickness decreases, the system moves towards the more symmetric state where $T$ and $B$ subsystems become identical. This will result in the change of the gap size and, for compressive strain, induce a semiconductor-metal transition. 

We show in the SI that at the $\Gamma$ point, the Hamiltonian can be separated into four families
\begin{equation}
H_{nm} = \Sigma + n PZ+m\left[ M+n K_CZ\right]L\,.
\label{eqn:Hnm}
\end{equation}
$\Sigma$ is on-site energy matrix; $P$ is the hopping matrix between $T$ and $B$ subsystems of the same sublattice ($C$ and $D$); $M$ describes hopping between $C_T$ and $D^T$; $K_C$ connects $C^T$ to $D^B$. $Z$ is a diagonal matrix with (1, 1, 1, $-1$) and $L$ is a diagonal matrix with ($-1$, 1, 1, 1). Finally, $m, n= \pm1$. States with $m = 1$ are anti-symmetric in $s$ orbitals for atoms on the same level ($T$ and $B$) and symmetric in other orbitals. For $n = 1$, the $T$ and $B$ atoms of the same syblattice are symmetric in $s$, $p_x$, and $p_y$ and anti-symmetric in $p_z$.

To understand how the deformation affects the gap size, it is helpful to consider two limiting cases: a completely flattened layer and a layer where the bonds connection $T$ and $B$ subsystems are perpendicular to them (maximum puckering). In the first case, the system becomes identical to graphene. Here, $p_z$ orbitals become orthogonal to the rest. In addition, $T/B$ symmetry is restored and $n = -1$, yielding the following energies at the $\Gamma$ point
\begin{equation}
E_\text{flat} = \Sigma_z+m M_z\,.
\end{equation}
$\Sigma_z$ is the energy arising from the overlap of the single sublattice ($C$ or $D$) $p_z$ orbitals and $M_z$ is the sum over all the $\pi$ bonds between the sublattices. Since the hopping element for $\pi$ bonds is negative, $M_z<0$. This means that the state where all the $p_z$ orbitals are aligned in the same direction have the energy $E = \Sigma_z+M_z$, smaller than the state where the sublattices are antisymmetric in $p_z$ ($E = \Sigma_z-M_z$). The reason for this is that the symmetric arrangements results in bonding, which is lower in energy than the anti-bonding antisymmetric arrangement.

Let us now move on to the maximum-puckering case. Here, neighboring $C^{T/B}$-$D^{B/T}$ atoms are aligned along the $\hat z$ axis. This means that symmetrically aligned neighboring $p_z$ orbitals form and anti-bonding $\sigma$ bond instead of the bonding $\pi$. Similarly, anti-symmetric neighbors form a $\sigma$ bond instead of the $\pi$ anti-bond. In fact, the general nature of $C^{T/B}$-$D^{B/T}$ interaction becomes more bonding for the anti-symmetric case and more anti-bonding for the symmetric case as one goes from a flat to a puckered system. If the $\sigma$ bond energy is substantially larger than that of the $\pi$ bond, puckering can actually cause the previously anti-bonding arrangement to become bonding and vice versa. In fact, according to our numerical calculations, the lowest conduction band is described by $H_{-1,-1}$, see Eq.~\eqref{eqn:Hnm}, corresponding to the fully symmetric $p_z$ orientation. On the other hand, the highest valence band has $n  = -1$, $m = 1$, which is anti-symmetric. Clearly, the ordering of the bands is opposite to what one finds in a flat layer. This means that layer compression leads to the gap reduction and an eventual band crossing.

\begin{figure}
\includegraphics[width = \columnwidth]{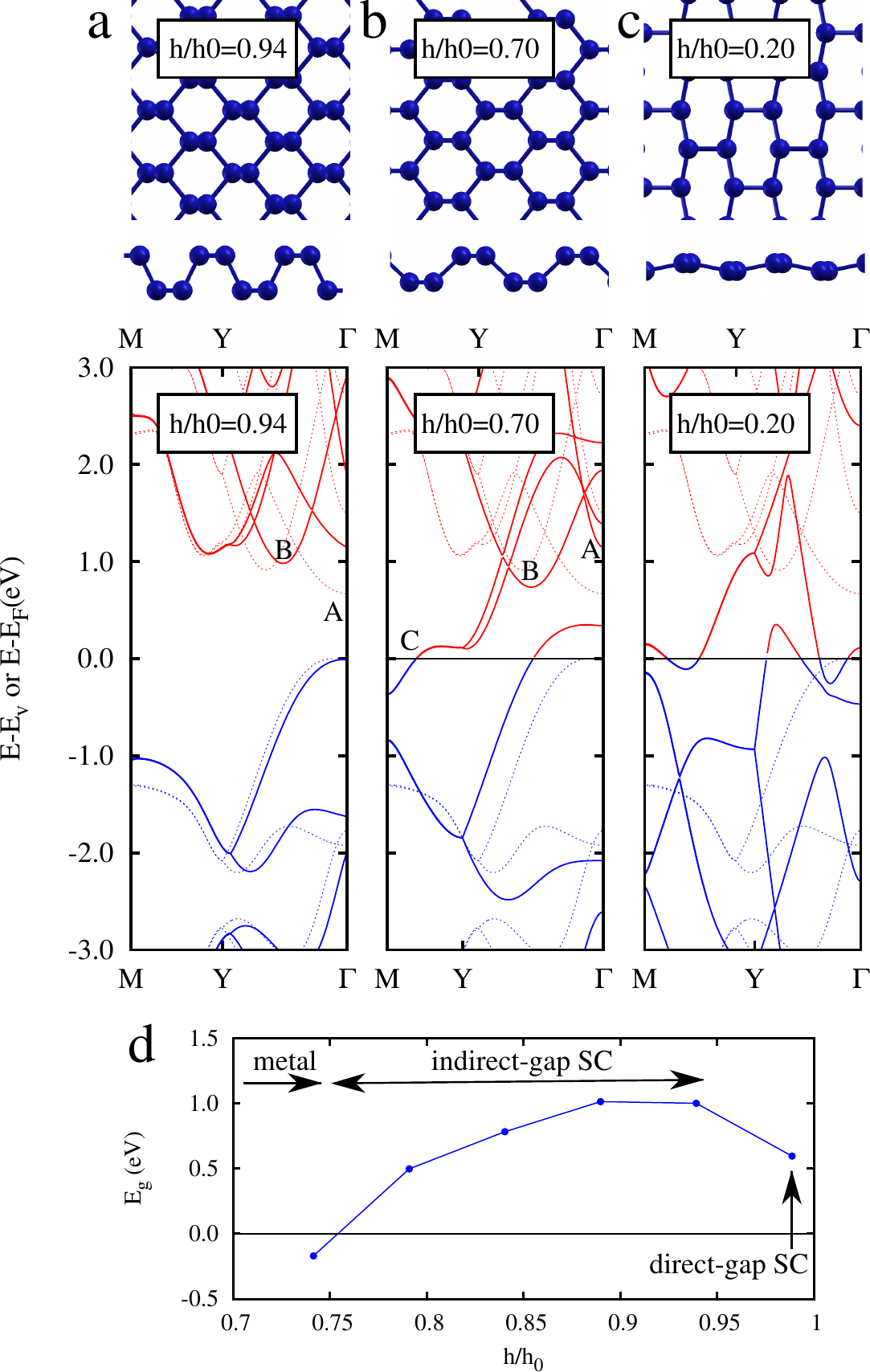}
\caption{
(a-c) Bandstructure of black phosphorus monolayer under uniaxial compression along the $z$ direction,
for three values of the imposed height $h/h_0$ (continuous line),
along with the bandstructure of the pristine material (dotted line).
The respective relaxed structure is also depicted in top and side view.
(d) Bandgap as a function of the height. 
The original layer thickness is 2$h_0$.
}
\label{bands-strain}
\end{figure}

To confirm this conclusion, we modeled the strained layers using density functional theory.
The monolayer unit cell and atomic positions were relaxed subject to the constraint $z=\pm h$
for all atoms. Compression ($h<h_0$, where $2h_0$ is the thickness of the free layer)
results in an in-plane expansion of the unit cell.
Until $h/h_0\sim 0.4$, the bonding structure of black phosphorus remains, 
and the structure of the strained material approaches that of 
a puckered graphene layer (Fig.~\ref{bands-strain}-a,b).
Below $h/h_0\sim 0.2$, however, there is a transformation into a square lattice (Fig.~\ref{bands-strain}-c).

With regard to the bandstructure, the valley at $\Gamma$, marked A in Fig.~\ref{bands-strain}-a,
first raises while the valley B becomes lower in energy. 
Hence, at $h/h_0=0.94$, the material is an indirect-gap semiconductor.
With further compression, a new valley appears at the Y point (marked C Fig.~\ref{bands-strain}-b).
For $h/h_0\sim 0.75$, this one eventually becomes as low as the valence band top near $\Gamma$,
marking the transition from indirect-bandgap semiconductor to metallic.
Figure~\ref{bands-strain}-d shows the variation of the bandgap energy, until a point 
where the conduction band minimum has descended below the valence band maximum.

In a narrow range of $h/h_0$ between $0.75$ and $0.70$, the material 
has a low density of states at the Fermi level and can be considered a semimetal.
Hoewever, different from graphene, there are electron and hole pockets 
in separate zones of the reciprocal space.
If compression is increased, the original valence bands and conduction bands finally cross,
as predicted, but the Dirac-like points are above the Fermi level.

Below $h/h_0\sim 0.2$ where the lattice is already nearly square, 
the resemblance of the bands with those of the original material is completely lost
(Fig.~\ref{bands-strain}-c).

\emph{Conclusions---}
Using \emph{ab initio} calculations, we have obtained the band structure of single-layer black phosphorus. The results show that this material is a direct-bandgap or nearly-direct band gap semiconductor with a stronly anisotropic dispersion in the vicinity of the gap. From the first principles calculation we also obtain the localized orbital composition of the bands arond the $\Gamma$ point which allows us to construct an effective Hamiltonian which describes the highest valence and the lowest conduction bands.

Based on the lattice structure of black phosphorus, we use a general tight-binding description to predict the closing of the gap with compression in the transverse direction. To support this prediction, we use DFT to show that upon moderate deformation, the system goes through a semiconductor-metal transition.  The energy ordering of the conduction band valleys change with strain
in such a way that it is possible to switch from nearly-direct bandgap semiconductor to indirect semiconductor, semimetal and metal with the compression along {\em only one direction}. Finally, under severe compression, the monolayer approaches a plane square lattice configuration. Such rich variety of eletronic and structural transformations make P-black an unique material for fundametal physics studies.

A.S.R. acknowledges DOE grant DE-FG02-08ER46512, ONR grant MURI N00014-09-1-1063. A.H.C.N. acknowledges NRF-CRP award ``Novel 2D materials with tailored properties: beyond graphene" (R-144-000-295-281). The first-principles calculations were carried out on the GRC high-performance computing facilities.


%

\end{document}